\documentclass{article}
\usepackage{graphics,color,array,dcolumn}
\usepackage{hyperref}
\usepackage{amsmath}
\usepackage{amssymb}

\def\GW{graviweak}
\def\sutwo{SU(2)}

\def\sufour{SU(4)}
\def\sutwol{\sutwo_L}
\def\sutwor{\sutwo_R}

\def\sothreeone{SO(3,1)}
\def\sothreeonec{SO(3,1,{\C})}

\def\sofourc{SO(4,{\C})}
\def\sofourcl{SO(4,{\C})_L}
\def\sofourcr{SO(4,{\C})_R}
\def\sosixc{SO(6,{\C})}

\def\sosevenc{SO(7,{\C})}
\def\soeightc{SO(8,{\C})}
\def\soten{SO(10)}

\def\sothirteenc{SO(13,{\C})}
\def\sothirteenone{SO(13,1)}

\def\sltwoc{SL(2,{\C})}

\def\cS{{\cal S}}

\def\det{{\rm det}}
\def\tr{{\rm tr}}

\newcommand\be{\begin{equation}}
\newcommand\ee{\end{equation}}
\newcommand\bea{\begin{eqnarray}}
\newcommand\eea{\end{eqnarray}}

\newcommand\psiL{\psi_L}
\newcommand\psiR{\psi_R}
\newcommand\PsiL{\Psi_L}

\newcommand\one{{\mathbf{1}}}
\newcommand\al{\alpha}
\renewcommand\b{\beta}
\newcommand\s{\sigma}

\newcommand\sh{\hat\sigma}
\renewcommand\t{\theta}
\newcommand\T{\Theta}
\newcommand\w{\omega}

\newcommand\e{\epsilon}
\newcommand\q{\ensuremath{{\scriptstyle\,\wedge\,}}}
\newcommand\m{\mu}
\newcommand\n{\nu}

\newcommand\Dt{{\tilde\Delta}}
\newcommand\D{\Delta}
\newcommand\de{\partial}
\newcommand\Ztwo{\ensuremath{{\mathbb Z}_2}}

\newcommand\RR[1]{\ensuremath{\mathbf{#1}}}
\newcommand\Rb[1]{\ensuremath{\mathbf{\overline{#1}}}}
\newcommand\C{\ensuremath{\mathbb C}}

\makeatletter
\renewcommand\section{\@startsection {section}{1}{\z@}%
                                   {-3.5ex \@plus -1ex \@minus -.2ex}%
                                   {2ex \@plus.2ex}%
                                   {\normalfont\large\bfseries}}
\makeatother

\begin{document}

\title{Gravi-Weak Unification}

\author{
{Fabrizio Nesti$^a$, Roberto Percacci$^b$}\\[.5em]
%{percacci@sissa.it}\\
\emph{$^a\,$Universit\`a dell'Aquila \& INFN, LNGS, I-67010, L'Aquila, Italy}\\
\emph{$^b\,$SISSA, via Beirut 4, I-34014 Trieste, Italy}\\
\emph{INFN, Sezione di Trieste, Italy}
}

\date{June 22, 2007}

\maketitle

\begin{abstract}
\noindent 
The coupling of chiral fermions to gravity makes use only of the
selfdual $\sutwo$ subalgebra of the (complexified) $\sothreeone$
algebra. It is possible to identify the antiselfdual subalgebra with
the $\sutwol$ isospin group that appears in the Standard Model, or
with its right-handed counterpart $\sutwor$ that appears in some
extensions.  Based on this observation, we describe a form of
unification of the gravitational and weak interactions.  We also
discuss models with fermions of both chiralities, the inclusion strong
interactions, and the way in which these unified models of
gravitational and gauge interactions avoid conflict with the
Coleman-Mandula theorem.
\end{abstract}
%\endinsert

\section{Approaches to unification}

In particle physics, the word ``unification'' is used in a narrow
sense to describe the following situation. One starts with two set of
phenomena described by gauge theories with gauge groups $G_1$ and
$G_2$.  A unified description of the two sets of phenomena is given by
a gauge theory with a gauge group $G$ containing $G_1$ and $G_2$ as
commuting subgroups.  In the symmetric phase of the unified theory,
the two sets of phenomena are indistinguishable.  The subgroups $G_1$
and $G_2$, and hence the distinction between the two sets of
phenomena, are selected by the vacuum expectation value (VEV) of an
``order parameter'', which is usually a multiplet of scalar fields.
The Standard Model (SM) and its grand unified extensions work this
way.  On the other hand, currently popular theories that claim to
provide a unification of gravity with the other interactions do not
fit into this general scheme.

It is possible to unify gravity and other Yang-Mills interactions in
the sense described above, if one allows the order parameter to be a
set of one-forms instead of scalars~\cite{so14}.  In order to motivate
this, let us begin by considering four one-forms
$\theta^m{}_\mu$ ($m=0,1,2,3$) transforming under ``internal'' global
Lorentz transformations $\t^m{}_\mu\to S^{-1 m}{}_n\t^n{}_\mu$,
which preserve the internal Minkowski metric $\eta$, and
linear coordinate transformations $\t^m{}_\mu
\to\t^m{}_\nu\Lambda^\nu{}_\mu$. 
Let us suppose that the dynamics of the theory is such that the
$\theta$ has a constant VEV
$\langle\t^m{}_\nu\rangle=\bar\t^m{}_\nu$ with
\be
\det\,\bar\t^m{}_\nu\not=0\,.
\ee
This VEV breaks the original invariances but preserves the global
``diagonal'' Lorentz subgroup defined by $\Lambda=\bar\t^{-1}S
\bar\t$.  Defining $\t=\bar\t+h$, the matrix $H=\eta h\bar\theta^{-1}$
(which has two covariant latin indices) transforms under the unbroken
group as $H\to S^T H S$. Therefore its symmetric and antisymmetric
parts, which have ten and six components respectively, are irreducible
representations of this unbroken group.  On the other hand, under an
infinitesimal internal Lorentz transformation $S=1+\epsilon$, the
antisymmetric part of $H$ gets shifted: $H\to H-\eta\epsilon$.  This
is the typical behavior of Goldstone bosons.  As a result, if the
action was invariant under the original transformations, the
antisymmetric components of $H$ would be massless.

Gravity in first order formulation is a gauged version of the
preceding theory, where the gauge field is a local Lorentz connection
$A_\mu{}^m{}_n$, the field $\t^m{}_\mu$ is the vierbein or soldering
form, defining a metric
\be
\label{eq:g1}
g_{\mu\nu}=\t^m{}_\mu\t^n{}_\nu\eta_{mn}\,,
\ee
and its covariant curl is the torsion tensor
\be
\label{eq:torsion1}
\T_\mu{}^m{}_\nu=\partial_\mu\t^m{}_\nu-\partial_\nu\t^m{}_\mu
+A_\mu{}^m{}_n\t^n{}_\nu-A_\nu{}^m{}_n\t^n{}_\mu\ .
\ee
As usual in gauge theories, the Goldstone bosons can be gauged away
(this corresponds to choosing the unitary gauge) and their kinetic
term generates a mass for the gauge fields in the ``broken''
directions.  For example, if the VEV is $\bar \t=M\one$, i.e.\ flat
space, the term
\be
\label{eq:TT}
\T^m_{\mu\nu} \T^{n\,\mu\nu}\eta_{mn}
\ee
generates a mass term of the form
$M^2(A_{mnp}-A_{pnm})(A^{mnp}-A^{pnm})$.  A similar phenomenon occurs
with the Palatini action. We have here a version of the Higgs
mechanism, where the six antisymmetric components of $H$ become the
six longitudinal components of the Lorentz connection.  The symmetric
components of $H$ cannot be eliminated in this way: after suitably
dealing with the diffeomorphism invariance, they correspond to the
graviton field.  It is possible to show that they are also Goldstone
bosons for a larger $GL(4)$ internal symmetry~\cite{libro}; this is
useful if one wants to discuss more general theories where the
connection is allowed not to be metric, but this will not be needed
for the purposes of this paper.  The main conclusion of this
discussion is that gravity is a gauge theory in the Higgs (``broken'')
phase, and that a ``symmetric'' phase would correspond to vanishing
$\bar \t$ and hence vanishing metric.\footnote{Because in the vierbein
formulation we are used to maintaining explicit invariance under local
Lorentz transformations, the statement that local Lorentz invariance
is broken spontaneously may cause some confusion.  In fact, gauge
symmetries are never broken: in the Higgs phenomenon it is the choice
of the unitary gauge that breaks the gauge invariance, but we are free
to choose any other gauge if we want to.  The physical meaning of
``broken gauge symmetry'' is that the gauge bosons are massive; this
is consistent with the absence of massless gauge particles connected
to the gravitational interactions.}

If one regards the soldering form as an order parameter, one sees
that gravity is unification-ready.  One way to achieve the unification
of gravity and other gauge interactions, in the strict sense defined
above, is to enlarge the internal spaces from four to $N$ dimensions.
For example it was proposed in~\cite{so14} that gravity could be
unified with an $SO(10)$ GUT in the group $\sothirteenone$, 
which contains the local Lorentz generators and the
$SO(10)$ generators as commuting subalgebras.  

In the present work we begin to discuss this unification mechanism
from the bottom up, starting from the weak interactions, rather than
postulating from the outset what the unified gauge group has to be.
This means identifying the local Lorentz gauge group of gravity and
the local isospin gauge group of the weak interactions as commuting
subgroups in a unifying group.  The direction is clearly indicated by
the quantum numbers of the SM fermions.  We shall see that if we
consider only chiral spinors of a single handedness, it is indeed
possible to achieve this goal with the group $\sofourc$.  Since this
group is not simple, one may object that this is not a true
unification.  However, we can define the representations and the
action in such a way that the theory is invariant under a discrete
\Ztwo\ group that interchanges the weak and gravitational
interactions.  In this sense one has a symmetric treatment of
these interactions, ready for extension to truly unified groups. 
The manifest asymmetry that we see in the real world can be attributed to
a single symmetry breaking phenomenon, occurring near
the Planck scale, which also gives mass to the gravitational
connection.

There are at least two different ways in which this scheme can be
enlarged to account for massive fermions.  One of them is based on the
notion of ``algebraic spinors'' and Clifford algebras.
In this case the unifying group of the gravitational and weak
interactions is $GL(4,{\C})$.  In the other case the unifying group is
$\sosevenc$, and further unification with the strong interactions
leads to the group $\sothirteenc$.

This paper is organized as follows.  In section~\ref{sec:chiral} we
will discuss a simplified world where only fermions of a given
chirality are present.  If the fermions are left-handed, they couple
only to the selfdual part of the Lorentz connection.  We will then
identify the antiselfdual component of the connection with the weak
gauge fields.  In section~\ref{sec:kin} we write an action functional
for the fermions that make sense also in a symmetric phase, and we
show how it reduces to the familiar form in the ``broken'' phase.  The
same is done for the gauge and gravitational degrees of freedom in
section~\ref{sec:gauge}. In section~\ref{sec:LR} we discuss more
realistic extensions where fermions of both chiralities are present,
and in section~\ref{sec:strong} we briefly discuss the resulting
scenarios for including strong interactions.  In
section~\ref{sec:coleman} we discuss the status of the global Lorentz
invariance and the way in which conflict with the Coleman-Mandula
theorem~\cite{colemanmandula} is avoided.
Section~\ref{sec:conclusions} contains concluding comments.

\section{A simple chiral world}
\label{sec:chiral}

The most striking property of the SM is that all fermions
are chiral (Weyl) spinors with respect to Lorentz transformations and
either singlets or Weyl spinors with respect to the weak gauge
group $\sutwol$.  In this section we shall consider a simplified world
where all fermions are massless and the weak singlets are absent.
Furthermore, we ignore strong interactions and consider only one weak
(left) doublet
$$
\psiL=\begin{pmatrix}\nu_L&e_L\cr\end{pmatrix}\ .
$$
All other doublets can be treated in the same way.

The central observation is the following.  Because these fields are
complex, they automatically carry a representation of the
\emph{complexified} Lorentz and weak groups.  The algebra of the
complexified Lorentz group $\sothreeonec$ consists of real linear
combinations of the rotation generators $L_j$, the boost generators
$K_j$ and their purely imaginary counterparts $iL_j$ and $iK_j$.  In
the case of the chiral fermion fields, 
the physical rotations and boosts are realized by the
generators $M^+_j=L_j+iK_j$ and $iM^+_j$ respectively, which together
generate a group $\sltwoc_+$. The generators $M^-_j=L_j-iK_j$ of
$\sothreeonec$ commute with the $M^+_j$ and can therefore be
identified with physical operations on spinors that have nothing to do
with Lorentz transformations.  In our simplified model we will
identify $\sltwoc_+$ with the Lorentz group, and the group generated
by the $M^-_j$ with the weak isospin gauge group $\sutwol$.  The
generators $iM^-_j$ are related to the weak isospin generators in the
same way as the boosts are related to the rotations, therefore we will
call them ``isoboosts'' and we will call the group $\sltwoc_-$
generated by $M^-_j$ and $iM^-_j$ the ``isolorentz group''.  It is
just the complexification of the isospin group.  The isoboosts are not
symmetries of the world and we will discuss their fate later on.  The
whole group $\sothreeonec\equiv\sofourc=\sltwoc_+\times \sltwoc_-$,
which contains both Lorentz and isolorentz transformations, will be
called the ``graviweak'' group.

To make this more explicit, we can arrange the components of $\psiL$
as a 2$\times$2 matrix whose columns are (left) chiral spinors under
the Lorentz group and whose rows are chiral spinors under the weak
group:
\be
\psiL=\begin{pmatrix}\nu_L^1&e_L^1\cr \nu_L^2&e_L^2\cr\end{pmatrix}
\ee
The Lorentz group acts on this matrix by multiplication from the left
and the isolorentz group acts by transposed multiplication from the
right (notice that this is called mathematically a left action).

The field $\psiL$ is therefore a bispinor: it carries a bi-index
$(A\alpha)$, the first acted upon by Lorentz and the second by
isolorentz transformations. It can also be seen as a vector of the
graviweak group:
\be
\label{eq:vanderwaerden}
\psiL^a=\sh^a_{A\al}\psiL^{A\al}\,,\qquad
\psiL^{A\al}=\sh_a^{A\al}\psiL^a\,,
\ee
where $a=1,2,3,4$, $\sh_a^{A\al}$ are the van der Waerden symbols:
$\sh_j=\s_j$ ($j=1,2,3$) are the Pauli matrices and $\sh_4=\one_4$.
The matrices $\sh^a_{A\al}$ are their ``inverses''.

In this notation the Lorentz and isolorentz groups act on $\psiL$
with the following generators:
\bea
\label{eq:Lorentzgen}
Lorentz: \qquad&& M_j^+=\sigma_j \otimes\one_2\qquad \equiv \s_j{}^{A}_B\,\delta^\al_\b\\
\label{eq:weakgen}
isolorentz: \qquad&& M_j^-=\one_2\otimes\sigma_j \qquad \equiv \delta^{A}_B\,\s_j{}^\al_\b\,.
\eea

While this terminology may sound unfamiliar, all that we have
described are the standard transformation properties of a massless
fermion doublet. However, reformulating things in this way 
is suggestive of a form of unification of the
gravitational and weak interactions, with the graviweak group as
unifying group.  The graviweak group is the direct product of the
Lorentz and isolorentz transformations and therefore it may seem that
no true unification has been achieved in this way.  However, it is
both mathematically and physically different to have a gauge theory of
the group $\sofourc$, with a single coupling constant, and of the
group $\sltwoc\times\sltwoc$, which in general has two.  In the
following we will investigate the possible existence of a symmetric
phase of the theory where the graviweak invariance is manifest, and of
a Higgs phase where it is broken.  The world as we know it will
obviously have to be identified with the latter.

We conclude this section observing that the preceding discussion can
be repeated word by word inverting the roles of left and right.  One
would then have a simplified chiral world where the only matter is
represented by massless, right-handed leptons
$$
\psiR=\begin{pmatrix}\nu_R&e_R\cr\end{pmatrix}
$$ 
In this case the group $\sothreeonec$ contains the Lorentz
transformations, generated by $M^-_j$ and $iM^-_j$, and the isolorentz
group contains the right-weak gauge transformations $M^+_j$.  Until we
address massive fermions, in the next two sections we will continue
discussing the case of left fermions only.

\section{The fermionic action}
\label{sec:kin}

To motivate what follows, we first recall the way in which one writes
the action of chiral fermions coupled to gravity (not unified with
other gauge fields), allowing for possibly degenerate soldering form
$\t^m{}_\mu$.  We begin by noting that the soldering form carries the
fundamental (vector) representation of the local Lorentz group, which
is isomorphic to the tensor product of a spinor and a conjugate spinor
representation.  We shall use indices $A,B\ldots$ for the spinor
representation and primed indices $A',B'\ldots$ for the conjugate
representation.  The isomorphism is given by the van der Waerden
symbols $\sh_m^{A'A}$.  It is sometimes convenient to think of the
soldering form as the bispinor-valued one-form
\be
\label{eq:sold1}
\t^{AA'}=\theta^{AA'}{}_\mu\, {\rm d}x^\m =\sh_m^{AA'}\theta^m{}_\mu\,{\rm d}x^\m\,.
\ee
The fermion action must contain a spinor, a conjugate spinor and one
derivative; the soldering form is the right object to covariantly
contract the indices carried by these objects.  The fermionic kinetic
term is written as
\be
\label{eq:fermion1}
\int\!{\rm d}^4x \,|\theta|\,\psi^{*\,A'}\theta_{A'A}^\mu D_\mu \psi^A\,.
\ee
Here $D_\mu\psi^A=\partial_\mu\psi^A+\omega_\mu{}^{A}{}_B\psi^B$ is
the Lorentz covariant derivative, $\theta_{A'A}^\mu$ is the inverse
soldering form and $|\t|$ its determinant.

In order to have an action that makes sense also for degenerate
soldering form one writes
\be
\label{eq:psikin1}
\int\psi^{*\,A'}D\psi^A \q\theta^{B'B}\q\theta^{C'C}\q\theta^{D'D} \epsilon_{(A'A)(B'B)(C'C)(D'D)} \,,
\ee
where
$$\epsilon_{(A'A)(B'B)(C'C)(D'D)}= 
\sh_{A'A}^m\sh_{B'B}^n\sh_{C'C}^r\sh_{D'D}^s\,\e_{mnrs}
%=
%i\e_{AD}\e_{BC}\bar\e_{A'C'}\bar\e_{B'D'} 
%- i\e_{AC}\e_{BD}\bar\e_{A'D'}\bar\e_{B'C'}
$$
is antisymmetric in the exchanges of the couples of indices $(A'A)$,
$(B'B)$, etc. Assuming that $\theta$ is nondegenerate,
eq.~(\ref{eq:psikin1}) reduces to~(\ref{eq:fermion1}).

\medskip

We want to generalize the previous discussion to write an action for
fermion fields coupled to weak gauge fields and gravity, in a way that
is invariant under graviweak transformations.  In the previous section
we learned that the fermions can be represented as graviweak vectors
$\psi^a$ and their conjugates $\psi^{*\bar a}$, while the graviweak
gauge field, in the representation carried by the fermions, is
$A_\mu{}^a{}_b$.  Since the graviweak group is gauged, we will need
first of all the corresponding covariant derivative
\be
\label{eq:covdersymm}
D_\mu \psiL^a=\partial_\mu \psiL^a+A_{\mu}{}^a{}_b \psiL^b\,.
\ee 
The gauge field $A_\m{}^a{}_b$ can be decomposed in
gravitational (selfdual) and weak (anti-selfdual) parts with
generators (\ref{eq:Lorentzgen}) and (\ref{eq:weakgen}). 
The (complex) gravitational gauge field will be denoted
$\w_\m^j$, while the antiselfdual gauge fields can be further split in
real part, the isospin gauge field $W_\m^j$, and imaginary part, the
isoboosts gauge field denoted $K_\m^j$:
\be
\label{eq:covdersplit}
D_\mu \psiL=\left[\partial_\mu + \w_\m^j M^{(+)}_j+(W_\m^j+iK_\m^j) M^{(-)}_j \right]\psiL\,.
\ee

We must then define the variable describing the gravitational field.
Following the argument that leads to~(\ref{eq:fermion1}) we need a
generalized soldering form $\t^{\bar ab}=\t^{\bar a b}{}_\mu\,{\rm
d}x^\m$ whose components can be used to invariantly contract the
fermion bilinear $\psiL^{*\bar a}\psiL^b$ to the covariant derivative
(\ref{eq:covdersplit}).

Before proceeding further, we shall introduce a little more notation.
We will use indices $m,n,\ldots$ for the vector representation of the
(Lorentz) group $\sltwoc_+$ and $u,v,\ldots$ for the vector
representation of the (isolorentz) group $\sltwoc_-$.  These
representations are again connected to the corresponding spinor
representations by the van der Waerden symbols:
\be
V^n\sh_n^{A'A}=V^{A'A}\,;\qquad
V^u\sh_u^{\alpha'\alpha}=V^{\alpha'\alpha}\,.
\ee
Using equation (\ref{eq:vanderwaerden}) we can convert the pair of
indices $\bar a a$ of the generalized soldering form to a pair of
bi-spinor indices: $\theta^{A'\alpha' A\alpha}= \sh_{\bar
a}^{A'\alpha'}\sh_{a}^{A\alpha}\,\theta^{\bar a a}$.  We can then
transform the pair of indices $A'A$ to a Lorentz vector index and the
pair $\alpha'\alpha$ to an isolorentz vector index.  In this way the
generalized soldering form can also be written as a one-form with a
Lorentz and an isolorentz vector index: $\theta^{m u}=
\sh^m_{A'A}\sh^u_{\alpha'\alpha}\theta^{A'\alpha' A\alpha}$.

We can now generalize (\ref{eq:psikin1}) and write a fermion kinetic
term as:
\be
\label{eq:psikinsymm}
\cS_{\psi}= \int\psiL^{*\,\bar a}D\psiL^a \q\t^{\bar
bb}\q\t^{\bar cc}\q\t^{\bar dd} \epsilon_{(\bar aa)(\bar
bb)(\bar cc)(\bar dd)} 
\ee
where $\epsilon_{(\bar aa)(\bar bb)(\bar cc)(\bar dd)}$ is an
\sofourc\ invariant tensor, totally antisymmetric under interchanges
of pairs of indices $(\bar aa)$, $(\bar bb)$, $(\bar cc)$, $(\bar
dd)$.  As we did above with $\theta^{m u}_\mu$, we can convert all the
indices on this tensor to pairs of vector indices.  In this notation,
we choose it as follows
\bea 
\epsilon_{(mu)(nv)(rw)(sz)}&=&
\e_{mnrs}(\eta_{uv}\eta_{wz}+\eta_{uw}\eta_{vz}+\eta_{uz}\eta_{vw})+\nonumber\\
&&{}+
(\eta_{mn}\eta_{rs}+\eta_{mr}\eta_{ns}+\eta_{ms}\eta_{nr})\e_{uvwz}\ .
\label{eq:epsilon}
\eea
If the combinations in the two lines on the r.h.s.\ had arbitrary
coefficients, this would still have the desired invariance and
antisymmetry properties; the particular combination~(\ref{eq:epsilon})
is the only one that is in addition invariant under the \Ztwo\ group
exchanging the gravitational and weak sectors.

The fermionic action~(\ref{eq:psikinsymm}) is built without using a
metric, and therefore it makes sense also in the symmetric phase.  As
long as $\t^{\bar aa}$ has zero VEV, it does not describe a standard
propagating (gaussian) theory, because it only contains interaction
terms.  The action is invariant under diffeomorphisms $x'(x)$ and \GW\
gauge transformations $S^a{}_b$, under which the fields transform as:
\be
\label{eq:transtheta}
\psiL\to S\,\psiL\,,\qquad \t_L{}_\mu \to \Lambda^\nu{}_\mu S^{\dagger}\t_L{}_\nu S\,,
\ee
where $\Lambda^\n{}_\m=\frac{dx^{\n}}{dx^{\prime\m}}$ and the graviweak indices
have been suppressed.

%\pagebreak[3]

The ordinary low energy world is described by a background geometry
that corresponds to Minkowski space and flat Lorentz and isolorentz
connections.  Both Lorentz and isoboost invariances must be broken at
some high scale, because neither of these gauge fields appears in the
low energy spectrum.  In the broken phase of the theory, at energies
above the electroweak scale, the fermions can be treated as massless
fields.  The background geometry must thus select a timelike direction
in the vector representations of the isolorentz algebra, while
providing a soldering of the vector representation of the Lorentz
algebra to the tangent spaces of spacetime.  Selecting the timelike isolorentz
direction along the fourth axis, the VEV corresponds to the choice
$\langle\theta^{m4}_\mu\rangle=M\delta^m_\mu$ and 
$\langle\theta^{mu}_\mu\rangle=0$ for $u=1,2,3$,
where $M$ is a mass parameter.  Translating back to the bispinor
indices, this corresponds to $\langle \t{}_\mu \rangle =
M\delta_\mu{}^m (\sh_m\otimes \one_2)$.  In order to describe in a
covariant fashion also non-flat geometries with weak curvature we will
consider backgrounds of the form:
\be
\label{eq:VEV}
\langle \t{}_\mu \rangle = M\, e^m{}_\mu \, (\sh_m\otimes \one_2)\,.
\ee
The dimensionless fields $e_\mu{}^m$ are now ordinary, real vierbeins
connecting the tangent space index $\mu$ to the internal vector index
$m$.  As is done usually, a metric can be built from the vierbein as
in equation~(\ref{eq:g1}).
%
%$$
%g_{\mu\nu} = e_\mu^m e_\nu^n\,\eta_{mn}
%$$ 
Notice that the VEV selects $\sltwoc_+$ for soldering with the
spacetime transformations, hence the signature of the resulting metric
is Minkowskian.

This VEV breaks the original group in the correct way to provide
global Lorentz and local weak (isospin) gauge invariance: the $(+)$
part of the $\sofourc$, corresponding to the Lorentz
generators~(\ref{eq:Lorentzgen}), and the imaginary part of the $(-)$
generators (the isoboosts) do not leave~(\ref{eq:VEV}) invariant, and
therefore are broken.  Thus, the only unbroken subgroup of the
original gauge group is the weak $\sutwol$.  In addition, the special
VEV $\theta^m_\mu=\delta^m_\mu$ is invariant under the global diagonal
$\sothreeone$ defined by
\be
\label{eq:diagonal}
S={\cal D}^{(\frac{1}{2},0)}(\Lambda)\ ,
\ee
where $S$ and $\Lambda$ are as in (\ref{eq:transtheta}).  This is the
usual Lorentz group and we shall discuss its role in
section~\ref{sec:coleman}.

At low energy the massive degrees of freedom can be ignored and the
covariant derivative~(\ref{eq:covdersplit}) reduces to
\be
\label{eq:covder}
{\cal D}_\mu \psiL=\big(\partial_\mu+\w_\mu{}^i(\sigma_i\otimes\one_2) +
W_{\mu}{}^i(\one_2\otimes \sigma_i)\big)\psiL\,,
\ee
where now $\w$ is the VEV of the spin connection constructed from
$e^m{}_\mu$.  It vanishes on a flat geometry, while in curved space it
coincides with the Levi-Civita connection in selfdual language. Apart
from strong interactions, this is the covariant derivative of
left-handed fermions in the SM coupled to gravity~\cite{changsoo}.
Correspondingly, when we insert the VEV (\ref{eq:VEV}) in the
fermionic action~(\ref{eq:psikinsymm}), this produces the ordinary
kinetic terms for an $\sutwol$ doublet of canonically normalized
spinors~$\PsiL=M^{3/2}\psiL$:
$$
%\label{eq:psikin}
\cS_{\psi}= \int\PsiL^{*\,A'\al'}\, {\cal D}\PsiL^{A\al} \sh_{A'A}^m
\delta_{\al'\al} \q e^n\q e^r\q e^s \epsilon_{mnrs}=\int\!{\rm
d}^4x\,|e|\, e^\mu_{m}\PsiL^{*\,\al} \sh^m {\cal D}_\mu \PsiL^\al\,,
$$
where we have suppressed the $\sltwoc$ Lorentz indices in the last
expression. Notice the emergence of the $\sutwo$ metric
$\delta_{\al'\al}$ from the antisymmetric tensor~(\ref{eq:epsilon}).

\section{Gauge and gravity dynamics} 
\label{sec:gauge}

We describe now the dynamics of the gauge and gravity degrees of
freedom.  As already mentioned, among the fluctuations, the Lorentz
and isoboost gauge fields should have high masses, as they are not
observed.  The isospin gauge fields on the other hand should be
effectively massless (until one introduces the mechanism breaking the
weak interactions) and in addition, there must be a massless graviton.
These degrees of freedom must emerge from a graviweak-invariant action
in the broken phase.  The action should be written in absence of a
metric, and this can be done in the first order formalism~\cite{fre}.
It turns out that the possible terms that one may write are quite
constrained by the gravi-weak symmetry.

The action will involve graviweak-covariant combinations of
derivatives of $\t$ and $A$: the generalized torsion two-form
%$\T^{\bar aa}=D\t^{\bar aa}$
%
\be
\T_{\mu\nu}^{\bar aa}=
\partial_\mu\t^{\bar aa}{}_\nu-
\partial_\nu\t^{\bar aa}{}_\mu+
\bar A_\m{}^{\bar a}{}_{\bar b}\t^{\bar b a}{}_\nu+
A_\m{}^a{}_b\t^{\bar a b}{}_\nu -
\bar A_\nu{}^{\bar a}{}_{\bar b}\t^{\bar b a}{}_\mu-
A_\n{}^a{}_b\t^{\bar a b}{}_\mu 
\ee
and the curvature two-form% $R^a{}_b=dA^a{}_b+A^a{}_c\q A^c{}_d$:
\bea
R_{\m\n}{}^{\bar aa\,\bar bb}\!\!\!&=&\!\!\!R_{\m\n}^{ab}\delta^{\bar a\bar
b}+\bar R_{\m\n}^{\bar a\bar b}\delta^{ab}\\
\label{eq:R}
R_{\m\n}{}^a{}_b\!\!\!&=&\!\!\!\de_\m A_\n{}^a{}_b-\de_\n A_\m{}^a{}_b+
A_\m{}^a{}_cA_\n{}^c{}_b - A_\n{}^a{}_cA_\m{}^c{}_b\,.
\eea

We discuss first the generalized Palatini action, which contains terms
linear in curvature and terms quadratic in torsion:
\bea
\label{eq:EHsymm}
\cS_{R1}\!\!\!\!\!&=&\!\!\!\!\!\frac{g_1}{16\pi}\int 
 R^{\bar aa \,\bar bb}\q \t^{\bar cc}\q \t^{\bar dd}\,\e_{(\bar aa)(\bar bb)(\bar cc)(\bar dd)}\\
\label{eq:torsionsymm}
\cS_\T\!\!\!\!\!&=&\!\!\!\!\!a_1\int\!\bigg[ 
  t_{\bar ee}^{\bar aa\,\bar bb}\,\T^{\bar ee} 
  +(t^2)\,\t^{\bar aa}\q \t^{\bar bb}
           \bigg]\q\t^{\bar cc}\q \t^{\bar dd}
 \e_{(\bar aa)(\bar bb)(\bar cc)(\bar dd)}
%+\!\T^{\bar aa}\q\T_{\bar aa}\,,\ \ 
\eea
where $t_{\bar ee}^{\bar aa\,\bar bb} $ are zero-form auxiliary fields
reproducing the components of $\T^{\bar ee}$.

In deriving the equations of motion (EOMs) it is convenient to split
the connection and curvature in selfdual and antiselfdual parts,
converting the graviweak indices $(\bar aa)$ to Lorentz and isolorentz
indices $(mu)$.  Then, the EOMs for the isolorentz (anti-selfdual)
connection are identically satisfied when one inserts the
VEV~(\ref{eq:VEV}), while the equation for the Lorentz (selfdual)
connection imply that the standard gravitational torsion vanishes:
\be
\T_{\mu\nu}^{m}\equiv
\partial_\mu e^{m}{}_\nu-\partial_\nu e^{m}{}_\mu+
\w_\m{}^{m}{}_{n}e^{n}{}_\n+\w_\m{}^{m}{}_{n}e^{n}{}_\m=0\,.
\ee
This fixes $\w_\m{}^m{}_n$ to be the Levi-Civita connection of
$e_\m^m$.  On the other hand the equation relative to $\t_\m^{mu}$
produces the Einstein equations for the background~$e_\m^m$.

One can understand better the dynamics of the gauge fields by
inserting the VEV~(\ref{eq:VEV}) in the action and neglecting
interaction terms.  The generalized actions~(\ref{eq:EHsymm})
and~(\ref{eq:torsionsymm}) become
\be
\cS_{R1}+\cS_\T \to  \int \!{\rm d}^4x\,\sqrt{g} \Big[
\frac{g_1}{16\pi} M^2 R+ 4a_1M^2
\left(\T_{\m\n}^m \T^{\m\n}_m + 10\, K^{j}_\m\,K_{j}^\m \right) 
\Big]\,.
\ee
Thus one should identify the Planck mass as $M_{PL}^2=g_1M^2$.  Then,
this shows that the isoboost gauge fields $K_\m^j$ acquire mass at the
Planck scale. As discussed in the introduction, also the
spin-connection $\w_\m^j$, which is contained in $\T_{\m\n}^m$ and
$R$, becomes massive.  This can be seen most clearly for the constant
background $e_\m^m=\delta_\m^m$; in curved backgrounds, it will
generate masses for the fluctuations of $\w$ around the Levi-Civita
connection of $e_\m^m$.  The W boson drops out of both terms because
it commutes with the VEV and thus it remains massless.

Next, one can introduce an action quadratic in graviweak curvature:
\bea
\label{eq:R2symm}
\cS_{R2}\!\!\!\!\!&=&\!\!\!\!\!\frac1{g_2^2}\int 
\!\bigg[r_{\bar ee\,\bar ff}^{\bar aa\,\bar bb}\,R^{\bar ee \,\bar ff}  
 + (r^2)\,\t^{\bar aa} \q\t^{\bar bb}\bigg]
\q\t^{\bar cc}\q \t^{\bar dd}\e_{(\bar aa)(\bar bb)(\bar cc)(\bar dd)}
%+\! R^{ab}\q R_{ab}\ \ \ \ 
\eea
Inserting the VEV~(\ref{eq:VEV}) and eliminating the $r_{\bar ee\,\bar
ff}^{\bar aa\,\bar bb}$ auxiliary fields, this action reduces to a
term quadratic in the gravitational curvature plus the standard
Yang-Mills actions for the weak gauge fields:
\be
\cS_{R2} \to \frac1{g_2^2}\int \!{\rm d}^4x \sqrt{g} \,\bigg(- R_{\mu\nu}^j R^{\mu\nu}_j
- W_{\mu\nu}^j W^{\mu\nu}_j-K_{\mu\nu}^{j} K^{\mu\nu}_{j}\bigg )\,.
\ee
Above the breaking scale, the gravi-weak symmetry manifests itself in
the equality of the coefficients of all the three terms, while below
the Planck scale the isoboosts and the spin connection are massive and
decoupled. Due to the vanishing torsion, the $R_{\m\n}^jR^{\m\n}_j$
term contains higher derivatives for the graviton; its effect is
however negligible relative to the Hilbert term at our energies: the
phenomenological limits on its strength are very
loose~\cite{Rsquared}.

In the broken phase the comparison between the strength of
gravitational and weak interactions has to be based on the effective
Newton constant $k^2/g_1 M^2$, where $k$ is the energy scale.  At
current available energies this effective coupling is extremely small,
while near the Planck scale it becomes comparable to the other
couplings. Since all the other gauge interactions seem to converge to
order-one couplings at a Grand Unification scale quite near the Planck
scale, this can be taken as a hint toward the complete unification
of all gauge and gravitational interactions. The present model where
the weak and gravitational interactions are treated on equal footing
is a step toward this direction.

\medskip

Let us discuss the low energy spectrum of this theory.  The \GW\
fermions $\psiL^a$ contain the SM fermions and remain massless in this
chiral world.  In a nonchiral world, they should receive mass at the
scale of electroweak breaking.  As discussed above, among the six
complex \GW\ gauge fields the three ``isoboosts'' and the six spin
connection fields correspond to broken generators and acquire a Planck
mass.  Only the three $W$ gauge fields remain massless in the broken
phase.  They should become massive at the lower energy scale of
$\sutwol$ breaking.

The generalized soldering field $\t^{\bar aa}_\m$ gives rise to
interesting structure in the broken phase.  The full field has 64 real
components that can be decomposed as follows:
\be 
\label{eq:components}
 \t_\mu = M e_\mu^{m}(\sh_m\otimes\one_2) +
 h_\mu^{m4}(\sh_m\otimes\one_2) + \D_\mu^{mj} (\sh_m\otimes\s_j)\,.
\ee
The first term on the r.h.s.\ is the background, and $h$ and $\D$ are
fluctuations. These be rewritten, using the VEV $e_\m^m$, as
\be
h_{\m\n}= e_\n^n h_\m^{n}\,,\qquad
\Dt_{\m\n}^j= e_\n^n \D_\m^{nj}\,.
\ee
The fluctuations around the background consist thus of the
16-components field $h_{\mu\nu}$, and three new (16 component) tensor
fields $\Dt_{\mu\nu}^j$, one for each value of the $\sutwol$ index
$j$.  Since nine generators of the graviweak group are broken
(corresponding to spin, boosts and isoboosts), nine of these fields
can be fixed by the choice of unitary gauge.  The natural choices are
the six antisymmetric components of $h_{\mu\nu}$ and the three traces
$\Dt_\mu^{\mu\,j}$.  In this gauge the remaining degrees of freedom
are the (ten) symmetric components of $h$, which after proper
treatment of the diffeomorphism invariance become the physical
graviton, and the (3~times~15) traceless components of $\Dt^j_{\m\n}$.
These latter fields can also be decomposed in antisymmetric (6
component) and symmetric traceless (9 components) fields, that are
copies of a (traceless) graviton. The emergence of these copies of
tensor fields was noticed in~\cite{chams_colored} where the gauge
group of gravity was extended to a generic $SL(2N,C)$.  In agreement
with that analysis, one can see that the antisymmetric component of
$\Dt$ does not get a kinetic term from the generalized EH
term~(\ref{eq:EHsymm}).

The symmetric part of $\Dt_{\m\n}^j$ represents new tensor particles
that constitute an $SU(2)$ isospin triplet, therefore they have
standard weak interactions and below the electroweak breaking scale
they will consist of one neutral and two charged components.  However
all the direct couplings of $\Dt_{\m\n}$ with matter are
Planck-suppressed, as for the graviton, because the common kinetic
term is normalized with~$M^2$.

Since these charged spin two particles are not observed at low energy,
one can suppose that they have escaped detection because their mass is
above the electroweak scale. While their mass could only be predicted in
a complete model, some useful observations can still be made.
First, one can write a $\sutwol$ gauge invariant mass like
$\tr(\Dt^j\Dt^j)$, which at first sight could be taken as large as the
Planck mass.  However, such a mass is actually part of the expansion
of the cosmological term around the background:\footnote{Since $\Dt^j$
are traceless, this mass term is equivalent to the standard
Pauli-Fierz mass for spin-two fields~\cite{PF}.}
\vspace*{-1ex}
\bea
\lefteqn{
\lambda \int \t^{\bar aa}\!\q \t^{\bar bb}\!\q\t^{\bar cc}\!\q \t^{\bar dd}
 \e_{(\bar aa)(\bar bb)(\bar cc)(\bar dd)}=}\qquad\qquad&\nonumber\\
%&\supset& 
%\lambda M^2 \int\!{\rm d}^4x\,|e|  e_\m^m
%e_\n^n\D_\r^{rj}\D_\s^{sj}\e_{\m\n\r\s}\e^{mnrs}\nonumber\\
&=& 
\lambda M^4 \int\!{\rm d}^4x |e|
+ \lambda M^2 \int\!{\rm d}^4x\,|e|  \, %\tr\Dt^j\tr \Dt^j-
\tr(\Dt^j\Dt^j)+\cdots\,.
\label{eq:PF}
\eea
It is therefore observationally constrained to be very small, because
it is connected with the cosmological constant.

On the other hand, a different mass term may arise from the coupling
with a Higgs field, that would give mass to $\Dt$ but not to the
graviton, as it was described in~\cite{chams_colored}.  By this
argument one may expect $\Dt$ to have mass in the weak range.
It is then interesting to note that if its mass were slightly above
the weak scale, for example $m_\Dt=300\,$GeV, this particle would
probably have escaped detection at LEP, being too heavy to be produced
(in pair) and having a small decay rate (mainly through Higgs bosons).
It should nevertheless be produced at LHC by standard Drell-Yan gauge
interactions  $qq\to W\to\Dt\Dt$ at energy above $2m_\Dt$,
and provide a nice signal of this theory.  It is also interesting to
speculate that depending on the model the lightest component of this
triplet (usually the neutral one) may be stable and only weakly
interacting, therefore being a candidate for dark matter. It would be
interesting to carry out such an analysis in a complete model
including the electroweak breaking sector.

\section{Models with both chiralities}
\label{sec:LR}

Even though in the SM left- and right-handed fermions occur in
different representations of the gauge group, there are many unified
models where at a more basic level the symmetry between left and right
is restored.  The minimal such models were based on the
Left-Right-symmetry~\cite{LR,LRbreaking} and the Pati-Salam partial
unification $\sutwol\times \sutwor\times\sufour$~\cite{PS}.  In these
models the hypercharge $U(1)$ group is enlarged to a group $\sutwor$
acting on the right-handed fermions in the same way as the weak
$\sutwol$ acts on the left-handed ones.  Then from the point of view
of the Lorentz and weak groups the fermions occur in the
representations (\RR2,\RR2) and (\Rb2,\RR2) of $\sltwoc\times\sutwol$ and
$\sltwoc\times\sutwor$ respectively (here we label representations by
their dimension).

\pagebreak[3]

This suggests a first possible model, where we take two of the toy
models considered in the previous section, with opposite chiralities,
and join them to construct a semirealistic model of gravi-weak
unification where fermions can have masses.  Because infinitesimal
Lorentz transformations are identified in one case with the $M^+_j$
generators and in the other case with $M^-_j$, we have to assume
independent Lorentz groups for the two chiralities.  Thus we start
from a group $\sofourcl\times\sofourcr$ where the left \GW\ group
$\sofourcl$ contains the left-Lorentz group $\sltwoc_L$ and the weak
gauge group $\sutwol$, while the right-\GW\ group $\sofourcr$ contains
the right-Lorentz group $\sltwoc_R$ and the internal gauge group
$\sutwor$.  In the low energy broken phase the physical Lorentz group
will have to be identified with the diagonal subgroup of
$\sltwoc_L\times\sltwoc_R$.  

In this model every massive fermion is realized by means of two fields
$\Psi_L$, $\Psi_R$: a complex $\sofourcl$-vector ($\sofourcr$-singlet)
and a $\sofourcr$-vector ($\sofourcl$-singlet).  When decomposed into
representations of their Lorentz and internal subgroups, they become
the desired left-handed doublet of $\sutwol$ plus right-handed
doublet of $\sutwor$.

The breaking of the two $\sofourc_{L,R}$ groups to the respective
$\sutwo_{L,R}$ subgroups follows the scheme described in the previous
sections, using separate generalized soldering forms
$\theta_{L\mu}^{\bar a a}$ and $\theta_{R\mu}^{\bar b b}$.  We assume
that both VEVs have the form~(\ref{eq:VEV}), so that they define the
same background metric.  Then, since the diffeomorphism group is
unique, equation~(\ref{eq:diagonal}) written separately for the left
and right transformations, will define a single residual global
$\sothreeone$ Lorentz group.\footnote{With different left and right
VEVs this would generally not be the case and Lorentz invariance may
be broken as in~\cite{nesti1}.}

The presence of two generalized soldering forms $\theta_L$ and
$\theta_R$ leads also to two graviton fields.  One of these is the
standard massless graviton that is massless thanks to diffeomorphism
invariance, the other graviton may be naturally massive as discussed
e.g.\ in~\cite{salam-strathdee,chams_complex,nesti1}.  Likewise, also
the fields $\Dt$ will be doubled, and thus one has two triplet tensor
fields $\Dt_L$, $\Dt_R$, of left and right isospin.

The appearance of two independent \GW\ groups may be somewhat
unpleasant.  It would be more elegant to have from the outset a single
copy of $\sltwoc$ to be identified with Lorentz transformations.
Because the representations are complex, we can think of the real
Lorentz group $\sltwoc$ as the complexification of the ``spin'' group
$\sutwo_{S}$ generated by $M^{+}_j$.  Then, we must look for
representations of $\sutwo_{S}\times\sutwol\times\sutwor$.  The
left-handed fermions are in the representation $(\RR2,\RR2,\RR1)$
while the right-handed fermions are in the $(\Rb2,\RR1,\RR2)$. 
% A look at the tables shows that
The (chiral spinor) representation \RR8 of $SO(7)$ decomposes precisely
into $(\RR2,\RR2,\RR1)\oplus(\RR2,\RR1,\RR2)$.  Thus a massive lepton
doublet can be neatly accommodated into a single spinor representation
of $\sosevenc$, by including both the left and the right (conjugated)
fermions in the same multiplet.

In this model the generalized soldering form needed to write the
fermionic action will have two spinor indices in the \RR8 instead of
two vector indices: $\theta^{\alpha'\alpha}_\mu$. Since in $\sosevenc$
one has $\Rb8\times\RR8=\RR{64}_{herm}\oplus\RR{64}_{antiherm}$, one can
take an hermitian soldering form consisting of 64 real fields (for
each spacetime index $\mu$). Similarly to the simple chiral world
described above, its VEV should be
$e_\m^m(\sh_m\otimes\one_2\otimes\one_2)$, that leaves unbroken the
three correct symmetries: the global Lorentz group plus the two local
gauge groups $\sutwol$, $\sutwor$.  The fluctuations of
$\t_\m^{\al'\al}$ around this VEV contains 220 fields: the graviton
$h_{\m\n}$ (10 components), two traceless tensor fields that are
triplets under the two isospin groups, $\Dt_{\m\n\,L}^j$,
$\Dt_{\m\n\,R}^j$ (45 components each), and a similar tensor field
that is triplet under both left and right isospin groups
$\Dt_{\m\n}^{{j_L}{j_R}}$ (120 components). All these new tensor
particles, charged under $\sutwol$ or $\sutwor$, should take mass at
the breaking of these symmetries.

Each of the two models described above has advantages and
disadvantages.  The model based on $\sofourc_L\times\sofourc_R$ has
only 12 complex generators (whereas the model based on $\sosevenc$ has
21 complex generators) and is therefore the minimal model that treats
gravity and the weak force in a symmetric fashion.  However, the gauge
group is not simple.  The model based on $\sosevenc$ has a simple
gauge group and each fermion doublet forms an irreducible
representation.  Although it is considerably larger than the previous
group, it is the minimal simple group that contains both Lorentz
transformations and the left and right isospin groups and such that
weak doublets are in one irreducible representation.

\medskip

We conclude this section by commenting on the possible origin of the
electroweak breaking. While this part of the analysis should be
carried on in a specific model including the hypercharge and the
strong interactions, we find it useful to sketch the various
possibilities.  As is well known, the Higgs fields of the SM is an
isospin doublet, Lorentz singlet.  Since this is not a \Ztwo-symmetric
representation, the Higgs should be accompanied by the corresponding
partner or embedded in a suitable larger representation.

From the point of view of group theory, the simplest possibility would
be the analog of a Dirac spinor.  This is the sum of a $\psi^A$ field,
that is a doublet of Lorentz, i.e.\ a Weyl spinor, and a $\phi^\al$
field that is a doublet of isolorentz, i.e.\ an isospin doublet.  The
latter could play the role of the Higgs, and its VEV would lead to the
correct breaking of isospin, without breaking the Lorentz invariance.
The VEV would effectively break the \Ztwo\ symmetry, and may follow
from a mechanism similar to the one leading to spontaneous breaking of
LR-parity~\cite{LRbreaking}. It is not clear however how a symmetry
between a scalar doublet and a spinor singlet could be made consistent
with the spin-statistics relation.

The $\sutwol$ symmetry could of course be broken also by the VEV of
other fields.  One example is a ``graviweak adjoint'' representation
$\Phi^a_b$, with VEV in the isolorentz sector only: $\bar\Phi^{(-)j}
(\one_2\otimes\s_j)$. This is equivalent to a ``triplet breaking''.
A similar breaking could be due to an additional VEV of the soldering
form, in the $\Dt_{\m\n}^{j}$ components (see~(\ref{eq:components})).
In general such a VEV would break the Lorentz symmetry: for example
$\Dt_{00}^{j}$ breaks the boosts transformations, that may be
marginally allowed, while $\Dt_{02}^j$ would break the rotations,
which is undesirable.  A special possibility would be if
$\Dt_{\m\n}^j$ were proportional to the background metric $g_{\m\n}$
for all $j$, i.e.\ $\langle \Dt_{\m\n}^j\rangle = g_{\m\n}\bar
\phi^j$, that amounts to have a VEV for the traces of $\Dt^j$.  In
this case it would not break Lorentz and it would behave just as a the
scalar triplet $\Phi^a_b$ just described.

A more promising possibility would be to introduce, in one of the
models presented above, a field transforming under both the left and
right $\sutwo$ weak groups, that would contain two left isospin
doublets~\cite{LR,LRbreaking}. Since the full electroweak breaking is
intimately linked also to the breaking of $B-L$ or unified color
groups, we shall defer a more complete analysis to a future work, and
turn to the inclusion of strong interactions.

\section{Including strong interactions}
\label{sec:strong}

The fermion quantum numbers strongly suggest that the color
$SU(3)$ group, together with the $U(1)$ group generated by $B-L$,
forms the real group $SU(4)\approx SO(6)$~\cite{PS}. 
Let us then discuss briefly the inclusion of this extended color group 
in the models discussed above.

In the $\sofourc_L\times\sofourc_R$ case, it is natural to try to
unify first the two groups in $\soeightc$.  Indeed, the two spinors
$(\RR4,\RR1)\oplus(\RR1,\RR4)$ can be obtained from the reduction of
the \RR8 (vector) of $\soeightc$.  Then, the fermions of one SM family
should be in the \RR8 (vector) of $\soeightc$ and in the \RR4 (chiral
spinor) of $SO(6)\approx SU(4)$.  Unfortunately with orthogonal groups
it is always impossible to obtain a product of vector and spinor
representations from the reduction of irreducible representations of a
larger group, and thus it seems that having decided in the first place
to introduce the gravi-weak orthogonal group $\sofourc$ we have
precluded the possibility of unifying it with the strong interactions
in a simple group.\footnote{The situation may change by looking for
embeddings in non-orthogonal groups.}

Instead, the $\sofourc_L\times\sofourc_R$ model points in the
direction of the framework described in~\cite{next} where, using
Clifford algebras, this unification is achieved in a different and
geometrical way. There the $\sufour$ groups are also duplicated as
$\sufour_L\times\sufour_R$ and, in the symmetric phase, the left and
right sectors have completely independent degrees of freedom, both in
the gauge and in the gravitational sectors.  In the broken phase, as
the graviweak groups are broken and reduced to the single Lorentz
group, also the two color groups should be broken to a diagonal
$\sufour$ and then to the color $SU(3)$.

\medskip

The approach based on $\sosevenc$ allows the complete unification in a
simple group but suffers from another problem.  Here a SM family is
contained in a representation $(\RR8,\RR4)$ of
$\sosevenc\times\sosixc$, and under the inclusion
$\sosevenc\times\sosixc\subset\sothirteenc$, one can indeed obtain the
$(\RR8,\RR4)$ from the reduction of the (spinor) $\RR{64} \to
(\RR8,\RR4) \oplus (\Rb8,\Rb4)$.  The breaking $\sosevenc\to\sltwoc
\times \sutwol\times\sutwor$, leads to the further decomposition
$(\RR8,\RR4)\to (\RR2,\RR2,\RR1,\RR4)\oplus(\Rb2,\RR1,\RR2,\RR4)$,
showing explicitly that this is exactly a family of the SM.  However
the decomposition of the $\RR{64}$ contains also a so-called mirror
family, $(\Rb8,\Rb4)\to
(\Rb2,\RR1,\RR2,\Rb4)\oplus(\RR2,\RR2,\RR1,\Rb4)$, that should be
disposed of somehow.  The problem of eliminating mirror fermions has
no simple solution, mainly because one can not give a gauge invariant
mass to the undesired chiral $\sutwo$ doublets~\cite{wilczek-zee}:
their mass would break the weak symmetry and thus it can not be higher
than the electroweak scale.

This approach is very close in spirit to the original $\sothirteenone$
unified theory~\cite{so14}, where the chiral spinor (\RR{64}) of
$\sothirteenone$ decomposes under $\sothreeone\times\soten$ into
$(\RR2,\RR{16})\oplus(\Rb2,\Rb{16})$, and the $(\Rb2,\Rb{16})$
representation consists of Weyl mirror fermions.

\section{Avoiding Coleman-Mandula}
\label{sec:coleman}

The Coleman-Mandula theorem~\cite{colemanmandula} states that under
certain natural hypotheses (which are very likely to hold in the real
world) the symmetry group of the S-matrix must be a direct product of
the Lorentz group and an internal symmetry.  It is usually interpreted
as a no-go theorem, forbidding a nontrivial mixing of spacetime and
internal symmetries.  Supersymmetry and certain quantum groups
famously manage to avoid the theorem: in these cases the symmetry is
not an ordinary Lie group, as assumed by the theorem.
The proposal for unification discussed in the previous sections is
based on ordinary Lie groups and thus superficially may seem to
violate the theorem.  We will discuss here why this is not the case.

To explain this point we have to discuss first the fate of the Lorentz
group in the proposed unified models.  It is important to distinguish
the local Lorentz transformations acting on the internal spaces, which
are gauge transformations and are present independently of the
background, from the global Lorentz transformations which are only
defined as the subgroup of diffeomorphisms $x'(x)$ that leave the
background (Minkowski) metric invariant.  We can then address the
question whether these transformations are broken or not. The
``unitary gauge'' choice (\ref{eq:VEV}), with $e_\m^{m} = \delta_\m^m$
breaks both the $\sltwoc_+$ local Lorentz and these global Lorentz
transformations.  However, as discussed in section~(\ref{sec:kin}),
the VEV is invariant when a global Lorentz transformation $\Lambda$ is
compensated by an internal transformation with parameter $S={\cal
D}^{(\frac{1}{2},0)}(\Lambda)$.  It is this global Lorentz symmetry group that
enters into a discussion of the Coleman-Mandula theorem.

One of the hypotheses of the Coleman-Mandula theorem is the existence
of a Minkowski metric.  Thus from the point of view of the theory
described above, this means that it can only apply to the ``broken''
phase, more precisely to the special case when the ground state is
flat space.  But we have shown that in the broken phase the residual
symmetries are precisely a global Lorentz symmetry and a local
internal symmetry.  This is in complete agreement with the 
Coleman-Mandula theorem.

The greater symmetry of the unified theory would only manifest itself in
the situation where the soldering form vanishes, which would
correspond to a symmetric ``topological'' phase of the theory; in that
phase there would be no metric on spacetime, let alone a Minkowski
metric, and the hypotheses of the Coleman-Mandula theorem would not
apply.
This argument applies also to other theories such as the one discussed
in~\cite{so14}. 

The main lesson to be drawn from this discussion is therefore that
there need not be a contradiction between the Coleman-Mandula theorem and
theories that mix internal and spacetime transformations:
the theorem does not forbid such a nontrivial mixing,
as long as it manifests itself only in a phase with no metric.

\section{Discussion and Conclusions}
\label{sec:conclusions}

\looseness=+1
There have been many attempts at unifying gravity with the other
interactions~\cite{Goenner}.  The one we discussed here generalizes in
the most straightforward way the philosophy and the procedures that
are believed to work, in particle physics, for the other interactions.
We have considered mainly the symmetry between the (left or right)
weak interactions and the gravitational ones, in a ``graviweak'' group
$\sofourc$.  The unification of spin and isospin transformations is
very natural in view of the quantum numbers of the fermion
fields.\footnote{It is well-known that spin and isospin
degrees of freedom can mix in solitonic solutions~\cite{jackiw-rebbi}.
In this work we have suggested that such mixing may occur at the level
of fundamental degrees of freedom.}  Since the group $\sofourc$ is not
simple, the gauge fields associated to its commuting factors
$\sltwoc_+$ and $\sltwoc_-$ could in general have different couplings,
and in this sense the two interactions would not be truly unified. In
analogy to what is done in left-right symmetric models, we have
postulated the existence of a discrete $\Ztwo$ symmetry exchanging the
two sectors, that would force this unification. This symmetry can be
seen as a remnant of the unification in a larger simple group, as
discussed in sections~\ref{sec:LR} and \ref{sec:strong}. Therefore we
have a unification of the gravitational and weak interactions in the
sense described in the introduction.

The \GW\ unification requires a generalized vierbein or soldering form
that naturally acts as an order parameter.  Its VEV defines the
gravitational background and at the same time selects the weak isospin
group as the only unbroken gauge group.
By using the soldering form as an order parameter we also avoid
the potential obstruction provided by the Coleman-Mandula theorem.  
The conditions of the theorem, in
particular the existence of a metric, are only satisfied in the broken
phase, and indeed in that regime the model predicts that the symmetry
of the theory is the product of spacetime and internal symmetries.

After the symmetry breaking, the extended soldering form gives rise,
in addition to the standard graviton, also to an isospin triplet
traceless tensor field. The gauge-invariant mass of this spin-two
triplet is connected with the cosmological constant and thus is
constrained to be small. On the other hand its mass may arise from the
mechanism of electroweak breaking, leading to the interesting
possibility that this particle, while having escaped detection up to
now, may be directly observed at LHC.

We have then discussed extensions of this scheme to include massive
fermions, and briefly analyzed also the inclusion of the strong
interactions.  At the moment there seem to be (at least) two possible
frameworks.

One is based on the use of $\sosevenc$ that unifies in a simple group
both the left and right isospin groups with the Lorentz group.  This
is the minimal simple group that can accommodate a massive fermion in
a single irreducible representation. After inclusion of the strong
interactions, this approach however suffers from the well known
problem of mirror fermions, because only vector-like fermions are
generated and the undesired chiral copy can not be given mass higher
than the weak scale~\cite{wilczek-zee}.  The situation is thus similar
to the one encountered in the approach originally proposed
in~\cite{so14}, where all the interactions are unified in a group
$\sothirteenone$ and gravity is separated by the VEV of a soldering
form. There the multiplets of the resulting $SO(10)$ gauge
interactions always appear in vector-like couples~$\RR{16}$+$\Rb{16}$.

A second scheme, closer in spirit to left-right theories, postulates
the duplication of the \GW\ group in left and right copies at the most
fundamental level, $\sofourc_L\times\sofourc_R$.  This approach does
not suffer from the problem of mirror fermions, but calls for such a
duplication also in the strong sector,~$SU(4)_L\times SU(4)_R$.  These
duplications point toward a more geometric framework, discussed in a
separate work~\cite{next}, that is based on the use of Clifford
algebras.

To conclude, the bottom-up approach that motivated the present work
has led to the result that one can successfully treat the weak and
gravitational interactions on equal footing. It also showed possible
ways to construct scenarios of complete unification including the
strong interactions.  The problems that arise are similar to the
typical ones that occur in grand unified theories; on the other hand,
contrary to common belief, there seems to be no fundamental obstacle
to the unification of gravity with other gauge interactions using the
familiar methods of particle physics.  We think that further
exploration of these scenarios will provide novel insight both at the
fundamental and phenomenological level.

\medskip

\section{Acknowledgements}
This work is partially supported by the MIUR grant for the Projects of
National Interest PRIN 2006 ``Astroparticle Physics'', and by the
European FP6 Network ``UniverseNet" MRTN-CT-2006-035863.  F.N. would
like to thank INFN and ICTP for kind hospitality and support during
the completion of this work.  We would like to thank S. Bertolini,
M. Fabbrichesi and G. Senjanovic for discussions.


\begin{thebibliography}{99}

\bibitem{so14}
  R.~Percacci,
  %``Spontaneous Soldering,''
  Phys.\ Lett.\  B {\bf 144} (1984) 37;
  %%CITATION = PHLTA,B144,37;%%
  Nucl.\ Phys.\  B {\bf 353} (1991) 271.
  %%CITATION = NUPHA,B353,271;%%

%\bibitem{gl4}    R.~Floreanini and R.~Percacci,
  %``Canonical Algebra Of Gl(4) Invariant Gravity,''
%  Class.\ Quant.\ Grav.\  {\bf 7} (1990) 975.
  %%CITATION = CQGRD,7,975;%%

\bibitem{libro}
 R.~Percacci,
  ``Geometry Of Nonlinear Field Theories'', World Scientific,  (1986)

\bibitem{colemanmandula}
  S.~R.~Coleman and J.~Mandula,
  %``ALL POSSIBLE SYMMETRIES OF THE S MATRIX,''
  Phys.\ Rev.\  {\bf 159}, 1251 (1967).
  %%CITATION = PHRVA,159,1251;%%

\bibitem{changsoo}
 L.~N.~Chang and C.~Soo,
  %``The Standard Model With Gravity Couplings,''
  Phys.\ Rev.\  D {\bf 53} (1996) 5682
  [arXiv:hep-th/9406188].
  %%CITATION = PHRVA,D53,5682;%%

\bibitem{fre} R. D'Auria, L. Castellani, P. Fr\'e,
  ``Supergravity and Superstrings'', Volume I,
Wordl Scientific, Singapore (1991)

\bibitem{Rsquared} 
 K.~S.~Stelle, %``Classical Gravity With Higher Derivatives,'' Gen.\
  Rel.\ Grav.\ {\bf 9} (1978) 353.  %%CITATION = GRGVA,9,353;%%

\bibitem{chams_colored} A.H. Chamseddine, \emph{Phys.\ Rev.}  \textbf{D70} (2004) 084006.

\bibitem{PF} M.~Fierz, W.~Pauli,
  %``On relativistic wave equations for particles of arbitrary spin in an
  %electromagnetic field,''
  \emph{Proc.\ Roy.\ Soc.} A {\bf 173}, 211 (1939).
  %%CITATION = PRSLA,A173,211;%%

\bibitem{LR}  R.~N.~Mohapatra and J.~C.~Pati,
  %``Left-Right Gauge Symmetry And An Isoconjugate Model Of CP Violation,''
  Phys.\ Rev.\  D {\bf 11} (1975) 566;
  %%CITATION = PHRVA,D11,566;%%

\bibitem{LRbreaking} 
  G.~Senjanovic and R.~N.~Mohapatra,
  %``Exact Left-Right Symmetry And Spontaneous Violation Of Parity,''
  Phys.\ Rev.\  D {\bf 12} (1975) 1502.
  %%CITATION = PHRVA,D12,1502;%%

\bibitem{PS}   J.C.~Pati and A.~Salam,
  %``Lepton Number As The Fourth Color,''
  \emph{Phys.\ Rev.\  D} {\bf 10} (1974) 275.

\bibitem{salam-strathdee} C.J. Isham, A. Salam, J. Strathdee, \emph{Lett. Nuovo Cim.} \textbf{5} (1972) 969.

\bibitem{chams_complex} A.H. Chamseddine, \emph{Phys.\ Rev.}  \textbf{D69} (2004) 024015.

\bibitem{nesti1}
  Z.~Berezhiani, D.~Comelli, F.~Nesti and L.~Pilo,
  %``Spontaneous Lorentz breaking and massive gravity,''
   \emph{Phys.\ Rev.\ Lett.}, to appear, {\tt [arXiv:hep-th/0703264]}.
  %%CITATION = HEP-TH/0703264;%%

\bibitem{next}
F. Nesti, {\tt hep-th/0706.3304}.

\bibitem{wilczek-zee}
 F.~Wilczek and A.~Zee,
  %``Families From Spinors,''
  Phys.\ Rev.\  D {\bf 25} (1982) 553.
  %%CITATION = PHRVA,D25,553;%%

\bibitem{Goenner}
  H.~F.~M.~Goenner,
  %``On the history of unified field theories,''
  Living Rev.\ Rel.\  {\bf 7} (2004) 2.
  %%CITATION = 00222,7,2;%%

\bibitem{jackiw-rebbi}
  R.~Jackiw and C.~Rebbi,
  %``Spin From Isospin In A Gauge Theory,''
  Phys.\ Rev.\ Lett.\  {\bf 36}, 1116 (1976).
  %%CITATION = PRLTA,36,1116;%%


\end{thebibliography}
\end{document}